\documentclass[lettersize,journal]{IEEEtran}
\usepackage{amsmath,amsfonts,amsthm,amssymb}
\usepackage{algorithmic}
\usepackage{algorithm}
\usepackage{array}
\usepackage[caption=false,font=normalsize,labelfont=sf,textfont=sf]{subfig}
\usepackage{textcomp}
\usepackage{stfloats}
\usepackage{url}
\usepackage{verbatim}
\usepackage{graphicx}
\usepackage{cite}
\usepackage{hyperref}
\usepackage{cleveref}

\usepackage{enumitem}

\newtheorem{theorem}{Theorem}

\newtheorem{assumption}{Assumption}

\usepackage{multirow}

\usepackage{pifont}

\usepackage{orcidlink}

\usepackage[table]{xcolor}
\definecolor{mycolor}{gray}{0.92}

\begin{document}

\title{Radiative-Structured Neural Operator \\ for Continuous Spectral Super-Resolution}

\author{Ziye Zhang\orcidlink{0000-0002-4889-2760}, 
Bin Pan\orcidlink{0000-0003-3063-1762},~\IEEEmembership{Member,~IEEE,} 
Zhenwei Shi\orcidlink{0000-0002-4772-3172},~\IEEEmembership{Member,~IEEE}

\thanks{
The work was supported by the National Key Research and Development Program of China under Grant 2022YFA1003800, the National Natural Science Foundation of China under the Grant 62571273, Grant 62125102 and Grant U24B20177.
\textit{(Corresponding author: Bin Pan.)}

Ziye Zhang and Bin Pan are with the School of Statistics and Data Science, LEBPS, KLMDASR and LPMC, Nankai University, Tianjin 300071, China (e-mail: \mbox{ziyezhang@mail.nankai.edu.cn}; \mbox{panbin@nankai.edu.cn}).

Zhenwei Shi is with the Image Processing Center, School of Astronautics, Beihang University, Beijing 100191, China (e-mail:  \mbox{shizhenwei@buaa.edu.cn}).
}

}

\markboth{IEEE TRANSACTIONS ON MULTIMEDIA}
{Shell \MakeLowercase{\textit{et al.}}: Radiative-Structured Neural Operator for Continuous Spectral Super-Resolution}


\maketitle

\begin{abstract}
Spectral super-resolution (SSR) aims to reconstruct hyperspectral images (HSIs) from multispectral observations, with broad applications in computer vision and remote sensing.
Deep learning-based methods have been widely used, but they often treat spectra as discrete vectors learned from data, rather than continuous curves constrained by physics principles, leading to unrealistic predictions and limited applicability.
To address this challenge, we propose the Radiative-Structured Neural Operator (RSNO), which learns a continuous mapping for spectral super-resolution while enforcing physical consistency under the radiative prior.
The proposed RSNO consists of three stages: upsampling, reconstruction, and refinement. 
In the upsampling stage, we leverage prior information to expand the input multispectral image, producing a physically plausible hyperspectral estimate. 
Subsequently, we adopt a neural operator backbone in the reconstruction stage to learn a continuous mapping across the spectral domain.
Finally, the refinement stage imposes a hard constraint on the output HSI to eliminate color distortion.
The upsampling and refinement stages are implemented via the proposed angular-consistent projection (ACP), which is derived from a non-convex optimization problem. 
Moreover, we theoretically demonstrated the optimality of ACP by null-space decomposition.
Various experiments validate the effectiveness of the proposed approach in both discrete and continuous spectral super-resolution.

\end{abstract}


\section{Introduction}

\IEEEPARstart{S}{pectral} super-resolution (SSR) aims to recover hyperspectral images (HSIs) from the given RGB or multispectral image (MSI).
SSR alleviates hardware limitations that prevent direct acquisition of HSIs with both high spatial and spectral resolution, thereby facilitating various downstream applications such as classification \cite{classification1, classification2}, unmixing \cite{unmixing1}, segmentation \cite{segmentation1}, and detection \cite{detection1, detection2}.
Nevertheless, this task is inherently ill-posed, as it requires inferring dense spectral signatures from limited and spectrally coarse observations.

Traditional SSR approaches mainly include model-based methods \cite{maloney1986evaluation}, regression-based methods \cite{nguyen2014early_regression, Wu2017early_regression2}, and sparse representation methods \cite{arad2016early_sparse_recovery, zhang2018early_cassi}.
These methods primarily rely on prior knowledge and physical constraints, which contribute to robustness and interpretability. 
However, they often fail to fully leverage large-scale data, which may lead to underfitting.

Advances in deep learning have led to a range of approaches based on deep neural networks.
Numerous methods employing convolutional neural networks (CNNs) \cite{shi2018hscnn+, miao2019lambdanet} and attention mechanisms \cite{MST++, xu2022nesr, li2023progressive} lead to notable improvements and outperform traditional approaches.
Deep learning-based methods can also be categorized according to their problem settings.
For example, fusion-based methods \cite{IR&ARF_fusion, Fang2024CS2DIPs, Zhu2026AFO} use paired low-resolution HSIs and high-resolution MSIs to extract and fuse spectral-spatial information, producing high-resolution hyperspectral images.
However, the practical application of fusion-based methods is hindered by the difficulty of obtaining registered HSI-MSI pairs.
Thus, single-image spectral super-resolution methods \cite{Wang2025WHANet, Physics-driven-DSFNet, li2023progressive, Kaiwei2023INR-HSISR, zhou2025hsact} have gained attention, as they require only the input MSI or RGB image at the inference stage.
Unsupervised spectral super-resolution has also been extensively studied to address the challenge of acquiring paired MSI-HSI data \cite{qu2023unmixing, unsupervised2025}. 

Deep learning-based approaches have achieved great success, but they typically formulate SSR as a finite-dimensional regression task learned purely from data.
In reality, spectra are continuous functions shaped by inherent physical priors, rather than discrete vectors produced by data-driven models.
Ignoring this fundamental property may result in physically inconsistent predictions, degraded generalization ability, and limited practical applicability.

To address this challenge, we propose the Radiative-Structured Neural Operator (RSNO), which models spectra in infinite-dimensional space and incorporates physically grounded spectral priors derived from atmospheric radiative transfer models.
Inspired by the two-stage paradigm \cite{shi2018hscnn+, miao2019lambdanet, Li2022DualStage}, the proposed framework comprises two standard stages, upsampling and reconstruction, and an additional refinement stage, as illustrated in \Cref{fig:overview}.
In the upsampling stage, the spectral response function (SRF) and radiative transfer priors are leveraged to generate an image with higher spectral resolution.
With the upsampled HSI, we employ a data-driven neural operator backbone to reconstruct a finer HSI estimate.
Since the neural operator is resolution-invariant and coordinate-based \cite{li2021fourier, kovachki2023neural}, we can compute the estimated HSI at arbitrary spectral scales.
Finally, we refine the HSI by projecting it onto the affine space defined by the SRF, thereby ensuring consistency with the original MSI after degradation.
The upsampling and refinement stages employ our proposed angular-consistent projection (ACP), a novel methodology derived from a non-convex affine space projection problem based on cosine similarity. 
It not only leverages physical priors to upsample the MSI but also enforces hard constraints to guarantee reconstruction consistency.
Additionally, the neural operator backbone adopts a U-shaped architecture equipped with spectral-aware convolution (SAC) layers to capture multi-scale spatial–spectral features.
Experiments on real-world datasets show that RSNO not only performs competitively in conventional finite-dimensional SSR, but also enables continuous spectral reconstruction, highlighting its effectiveness and flexibility in practical applications.

The main contributions of this article are summarized below:
\begin{itemize}

    \item We formulate spectral super-resolution as a continuous regression problem in infinite-dimensional space, supporting arbitrary-scale spectral super-resolution;

    \item We incorporate atmospheric radiative transfer priors into the data-driven framework, enabling radiative-structured and physically coherent spectral reconstruction;

    \item We derive the ACP method from a non-convex affine space projection problem based on cosine similarity, and theoretically demonstrate its optimality by null-space decomposition.
    
\end{itemize}

\section{Related work}

\subsection{Continuous Spectral Super-Resolution}

Continuous spectral super-resolution aims at recovering spectral curves at arbitrary resolutions.
Xu et al. \cite{xu2022nesr} first introduced a framework that reconstructs spectral images with an arbitrary number of bands and demonstrated superior performance over discrete methods in fixed-band reconstruction tasks.
Beyond this, they further upgraded the neural attention mapping module in their subsequent work \cite{xu2025nessr}, thereby enabling continuous reconstruction in both the spatial and spectral domains with state-of-the-art performance among discrete baselines.
Besides, Chen et al. \cite{Chen2023SpectralINR} and Li et al. \cite{Li2025MGIR} studied continuous HSI reconstruction within the Coded Aperture Snapshot Spectral Imaging (CASSI) framework.
Additionally, Zhou et al. \cite{Zhou2026ASSR_FTD} proposed an efficient method for arbitrary scale spatial-spectral super-resolution based on functional tensor decomposition.

Many continuous approaches are built upon implicit neural representations (INRs) \cite{chen2021liif}, which implicitly represent natural and complex images continuously and have inspired a wide range of arbitrary-scale super-resolution methods in computer vision \cite{Zhao2024ASSR, Nasir2026ImplicitINR}.

\subsection{Physics-Informed Spectral Super-Resolution}

While data-driven methods have achieved strong performance in spectral super-resolution, their lack of explicit physical constraints has motivated the development of physics-informed approaches \cite{lin2020physically, Guo2023StableHSISR, Liu2022PDASS, Huo2024recover, Wu2024HPRN, nullspace2025}. 
These methods seek to integrate physical models or domain-specific priors into learning-based frameworks to enhance reconstruction fidelity and interpretability. 
For example, Wu et al. \cite{Wu2024HPRN} incorporated comprehensive semantic priors to regularize and optimize the solution space of SSR;
Liu et al.~\cite{Liu2022PDASS} embedded a spectral unmixing model into the reconstruction framework and explicitly accounted for solar radiation and atmospheric absorption effects.
In addition, several works \cite{lin2020physically, Huo2024recover, unsupervised2025} approached SSR from the perspective of a linear inverse problem and derived null-space priors to constrain the solution space, realizing high-fidelity reconstruction.

\subsection{Operator Learning}
\label{sec:operator_learning}

Operator learning is an emerging framework in machine learning that extends the learning task from finite-dimensional vector mappings to mappings between infinite-dimensional function spaces.
Representative architectures include DeepONet \cite{lu2021learning} and neural operator \cite{li2021fourier, kovachki2023neural, rahman2023uno}.

The neural operator is formulated as follows.
Let $a:D \to \mathbb{R}^{d_a}$ denote the input function defined on a bounded domain $D \subset \mathbb{R}^{d}$, $u:D' \to \mathbb{R}^{d_u}$ denote the output function defined on $D' \subset \mathbb{R}^{d'}$. 
Neural operators $\mathcal{G}_\theta$ learn the non-linear mapping $\mathcal{G}^{\dag}$ between two infinite-dimensional spaces by using a finite collection of observations of input-output pairs from this mapping, which is,
\begin{equation}
    \mathcal{G}_\theta: \mathcal{A} \to \mathcal{U}, \quad \theta \in \Theta,
\end{equation}
where $\Theta$ is the parameter space, $\mathcal{A}=\mathcal{A}(D,\mathbb{R}^{d_a})$ and $\mathcal{U}=\mathcal{U}(D,\mathbb{R}^{d_u})$ are separable Banach spaces of function taking values in $\mathbb{R}^{d_a}$ and $\mathbb{R}^{d_u}$ respectively. Normally, neural operators consist of three components: 
\begin{enumerate}
    \item \textbf{Lifting:} Lift the input function $\{a:D \to \mathbb{R}^{d_a} \}$ to the first hidden representation $\{v_0:D \to \mathbb{R}^{d_{v_0}} \}$ via a point-wise function $\mathbb{R}^{d_a} \to \mathbb{R}^{d_{v_0}}$;
    
    \item \textbf{Iterative Kernel Integration:} Iteratively update the hidden representation
    \begin{equation}
    \label{eq:iterative_update}
        v_{t+1} = \sigma \left( W_tv_{t}+  \mathcal{K}(a; \phi)v_{t}  \right), \quad \text{for } t = 0,1,...,T-1,
    \end{equation}
    where $\mathcal{K}$ is the kernel integral operator parameterized by $\phi$, $W_t: \mathbb{R}^{d_{v_t}} \to \mathbb{R}^{d_{v_{t+1}}}$ represents the linear transformation, and $\sigma: \mathbb{R} \to \mathbb{R}$ is the activation function;
    
    \item \textbf{Projection:} Map the last hidden representation $\{v_T:D \to \mathbb{R}^{d_{v_T}} \}$ to the output function $\{u:D \to \mathbb{R}^{d_{u}} \}$ via a point-wise function $\mathbb{R}^{d_{v_T}} \to \mathbb{R}^{d_{u}}$.
\end{enumerate}

Neural operators have a wide range of mathematical and scientific applications.
Besides, neural operators have also been applied to super-resolution tasks in computer vision and remote sensing \cite{wei2023srno, zhang2024stno, pansharpening_NO, Zhu2026AFO}. 
Their inherent resolution-invariance property enables continuous super-resolution, making them particularly well-suited for these applications.

\subsection{Atmospheric Radiative Transfer Model}
\label{sec:art}

Atmospheric radiative transfer models are crucial components of many scientific disciplines. 
Generally, these models aim to estimate atmospheric effects on the solar radiation under specific circumstances, such as vertical profiles, seasonal variations, and geographical latitude. 
Examples of atmospheric radiative transfer models mainly include MODerate resolution atmospheric TRANsmission (MODTRAN) \cite{berk1999modtran4}, Santa Barbara Disort Atmospheric Radiative Transfer (SBDART) \cite{ricchiazzi1998sbdart}, and the Simple Model of the Atmospheric Radiative Transfer of Sunshine (SMARTS) \cite{gueymard1995smarts2, gueymard2019smarts}. In this article, the radiative spectral prior is predicted by the SMARTS.

\section{Methodology}
\label{sec:methodology}

\begin{figure*}[!t]
    \centering
    \includegraphics[width=0.99\linewidth]{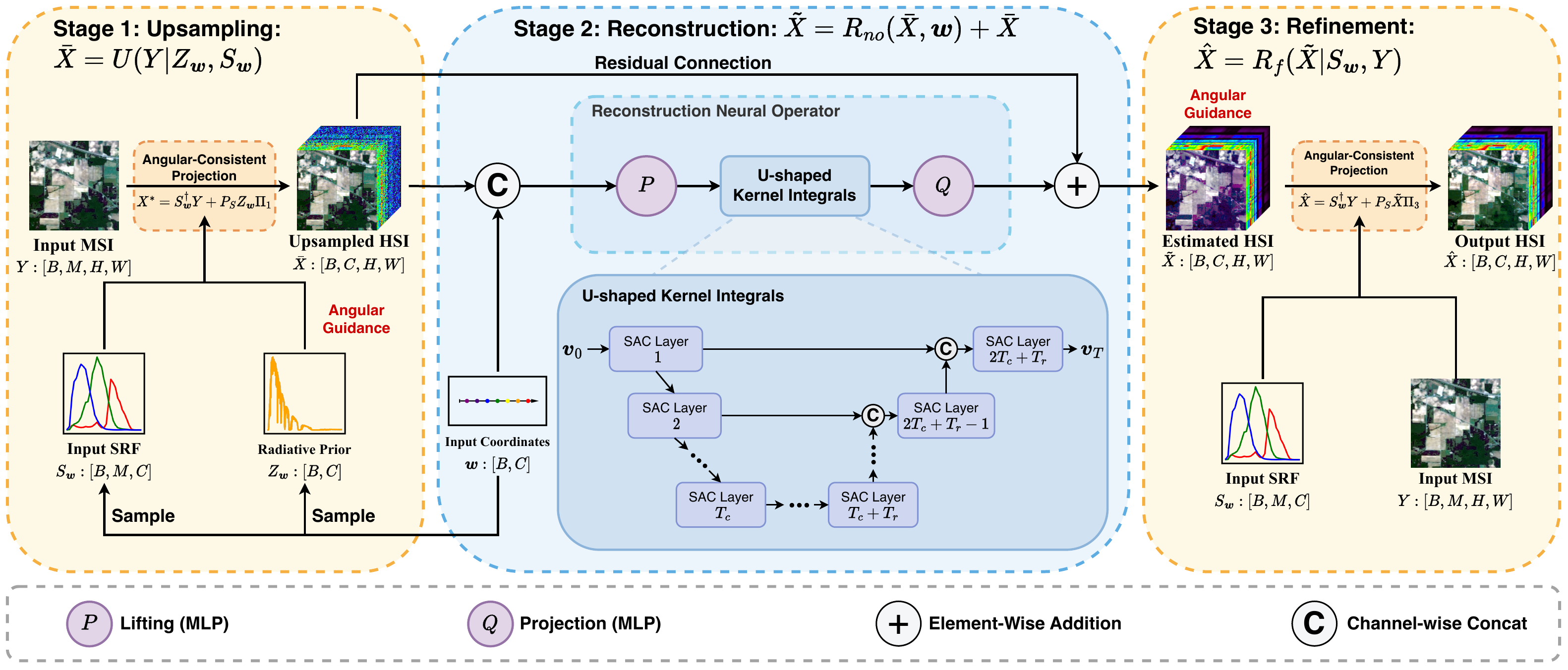}
    \caption{
    Overall framework of the proposed method. 
    The input MSI undergoes three main stages to output the final HSI $\hat{X}$: (1) \textbf{Upsampling}, where we expand the input MSI to the upsampled HSI, (2) \textbf{Reconstruction}, where we employ the neural operator to transform the upsampled HSI to a finer HSI estimate, and (3) \textbf{Refinement}, where we impose a hard constraint on the final result to enforce zero reconstruction error and eliminate color distortion. 
    }
    
    \label{fig:overview}
\end{figure*}

In this section, we present the RSNO methodology. 
We begin by formulating the spectral reconstruction problem and outlining the motivation and overall framework of our approach in \Cref{sec:problem_formulation}. 
We then introduce two key components of the framework: the proposed angular-consistent projection method and the U-shaped reconstruction neural operator with spectral-aware kernel integrals. 
\Cref{sec:acp} discusses the ACP method and its theoretical optimality, while \Cref{sec:no} provides a detailed description of the neural operator backbone.

\subsection{Overall Framework}
\label{sec:problem_formulation}

Let $X \in \mathbb{R}^{C \times N}$ denote the HSI, $Y \in \mathbb{R}^{M \times N}$ the paired RGB or multispectral image, and $S \in \mathbb{R}^{M \times C}$ the SRF, where $N=HW$ is the number of pixels, and $C$ and $M$ are the number of channels with $C>M$.  
Generally, SRFs contain enough information for each spectral band, so we assume that all SRF matrices are of full row rank.
The forward degradation process from the HSI $X$ to the MSI $Y$ can be mathematically formulated in a linear inverse problem, 
\begin{equation}
\label{eq:linear_inverse_problem}
    Y=SX+\mathcal{E},
\end{equation}
where $\mathcal{E} \in \mathbb{R}^{M \times N}$ is the pixel-wise noise. 
To statistically fit the model, we omit the noise and simplify it to the following underdetermined linear matrix equation:
\begin{equation}
\label{eq:Y=SX}
    Y=SX.
\end{equation}
Thus, the spectral reconstruction problem can be formally formulated as solving \eqref{eq:Y=SX}, where, given an RGB or multispectral image $Y$ and its corresponding spectral response function $S$, the objective is to recover a plausible solution $\hat{X}$ to \eqref{eq:Y=SX}.

To solve this ill-posed problem, a widely used deep learning-based approach is the two-stage reconstruction paradigm, where the input is first upsampled to the target resolution, followed by a reconstruction neural network that restores missing details and corrects estimation errors \cite{shi2018hscnn+, miao2019lambdanet, Li2022DualStage}. 
This paradigm can be formulated as  
\begin{equation}
    \label{eq:upsample_reconstruction}
    \hat{X} = R(U(Y)) + U(Y),
\end{equation}  
where $ U(\cdot) $ denotes the upsampling operation that projects the input $Y$ into a higher-dimensional target space, and $R(\cdot)$ is a learnable data-driven reconstruction module, such as CNN-based neural networks. 
A residual connection is employed to facilitate learning and enhance reconstruction quality.

However, this paradigm faces inherent limitations when applied to spectral reconstruction.
For the upsampler $U(\cdot)$, commonly used techniques such as bilinear interpolation and transposed convolution are less effective in spectral super-resolution, as the upsampling ratio is substantially higher than in typical spatial super-resolution tasks.
Additionally, this paradigm is entirely data-driven, lacking integration of physical processes.
Besides, most data-driven reconstruction modules rely on deep neural networks that learn mappings between finite-dimensional spaces, neglecting the continuity of spectral signals, thereby limiting both performance and applicability.

Hence, we adopt a similar yet more effective three-stage design, as illustrated in \Cref{fig:overview}.
Given the MSI $Y$, wavelength coordinates $\boldsymbol{w}$, and the SRF $S_{\boldsymbol{w}}$ discretized at $\boldsymbol{w}$, we sequentially apply the following three stages to obtain the final HSI estimate.

\begin{enumerate}
    \item \textbf{Upsampling:} In the first upsampling stage, we apply the proposed ACP method to obtain the initial upsampled HSI, which can be formulated as
    \begin{equation}
        \label{eq:stage1}
        \bar{X} = U(Y|Z_{\boldsymbol{w}}, S_{\boldsymbol{w}})=P_{\text{ACP}}(Z_{\boldsymbol{w}}, S_{\boldsymbol{w}}, Y),
    \end{equation}
    where $U(\cdot | Z_{\boldsymbol{w}}, S_{\boldsymbol{w}})$ is the conditional upsampler guided by $Z_{\boldsymbol{w}}$ and $S_{\boldsymbol{w}}$, $P_{\text{ACP}}(\cdot, \cdot, \cdot)$ is our proposed angular-consistent projector with given pixel-wise prior $Z_{\boldsymbol{w}} \in \mathbb{R}^{C \times N}$ as the angular guidance, SRF $S_{\boldsymbol{w}}$ and MSI $Y$, which are all discretized at $\boldsymbol{w}$.

    \item \textbf{Reconstruction:} In the next reconstruction stage, we employ a resolution-invariant neural operator with a U-shaped structure to learn the mapping between infinite-dimensional spaces of upsampled and target HSIs, which can be formulated as 
    \begin{equation}
        \label{eq:stage2}
        \tilde{X} = R_{no}(\bar{X}, \boldsymbol{w}) + \bar{X},
    \end{equation}
    where $\tilde{X}$ is the reconstructed HSI, $R_{no}(\cdot,\cdot)$ is the neural operator that takes the functional value and the coordinate as inputs, and a residual connection is employed similar to \Cref{eq:upsample_reconstruction}.

    \item \textbf{Refinement:} In the final refinement stage, we project the output to the solution space of \eqref{eq:Y=SX}, which imposes a hard constraint on our result. This stage can be formulated as
    \begin{equation}
        \label{eq:stage3}
        \hat{X} = R_{f}(\tilde{X} |S_{\boldsymbol{w}}, Y) = P_{\text{ACP}}(\tilde{X}, S_{\boldsymbol{w}}, Y),
    \end{equation}
    where $\hat{X}$ is the output HSI, $R_{f}(\cdot |S_{\boldsymbol{w}}, Y)$ is the refinement operator.
    Although the $R_{f}(\cdot |S_{\boldsymbol{w}}, Y)$ employs the same ACP procedure as the upsampler in the first stage, its objective is fundamentally different.
    The upsampling stage aims to produce a physically plausible upsampled HSI. 
    In contrast, the refinement stage projects the output onto the solution space of \eqref{eq:Y=SX}, thereby enforcing the corresponding hard constraint.
\end{enumerate}

Our framework successfully addresses the limitations of the traditional two-stage paradigm. 
First, the ACP method embeds physical priors into the data-driven framework through pixel-wise guidance. 
It also refines the output by solving a constrained optimization problem, thereby enforcing the hard constraint defined in \eqref{eq:Y=SX}.
Moreover, previous studies have shown that neural operators can accommodate inputs defined on meshes with different discretizations \cite{li2021fourier, kovachki2023neural}, so we employ a neural operator backbone to address the inherent limitation of conventional neural networks that can only model discrete spectral vectors. 

Within the proposed framework, the angular-consistent projection method and the reconstruction neural operator serve as two core components, which are detailed in \Cref{sec:acp} and \Cref{sec:no}, respectively.

\subsection{Angular-Consistent Projection}
\label{sec:acp}

In this section, we propose a novel method termed \textit{Angular-Consistent Projection}, based on the formulation in \Cref{sec:problem_formulation}. 
We seek a physically plausible HSI estimate that remains close to the prior $Z$ while satisfying the constraint in \eqref{eq:Y=SX}.
To achieve this, we can consider the following affine space projection problem:
\begin{align}
\label{eq:affine_projection_problem}
\min_X& \quad \text{SAM}(X, Z) \\
\text{s.t.}& \quad Y=SX, \nonumber
\end{align}
where $\text{SAM}(\cdot, \cdot)$ denotes the Spectral Angle Mapper (SAM), a widely used metric for measuring spectral similarity between two HSIs. It is defined as
\begin{equation}
\label{eq:sam}
\text{SAM}(X, Z) = \frac{1}{N} \sum_{n} \arccos \left( \frac{X_{\cdot n}^\top Z_{\cdot n}}{\Vert X_{\cdot n}\Vert \Vert Z_{\cdot n} \Vert} \right),
\end{equation}
where $\Vert \cdot \Vert$ is the $L2$ norm, $X_{\cdot n}$ denote the $n$th column vector of $X$, that is, the discretized spectral curve of $X$ at pixel $n$. 
Equivalently, it suffices to solve
\begin{align}
\label{eq:simplified_problem}
\max& \quad \frac{X_{\cdot n}^\top Z_{\cdot n}}{\Vert X_{\cdot n}\Vert \Vert Z_{\cdot n} \Vert} \\
\text{s.t.}& \quad Y_{\cdot n} = S X_{\cdot n}, \nonumber
\end{align}
for $n=1,2,...,N$. Therefore, our goal is to find an appropriate spectral estimate $X_{\cdot n}$ within the affine space $\mathcal{A}_n(S, Y_{\cdot n}) = \{ \boldsymbol{v} \in \mathbb{R}^C: S\boldsymbol{v} = Y_{\cdot n} \}$ that minimizes the spectral angle (or equivalently maximizes the cosine similarity) with the prior $Z_{\cdot n}$, for each pixel $n = 1, 2, \dots, N$.

However, since problem \eqref{eq:simplified_problem} is non-convex, the existence of a non-trivial solution is not guaranteed. Under the following assumption, we establish the existence of a non-trivial global optimum.

\begin{assumption}
\label{asp:problem}
    For all $n=1,2,...,N$, the SRF matrix $S \in \mathbb{R}^{M \times C}$, prior $Z_{\cdot n} \in \mathbb{R}^C$, and MSI pixel value $Y_{\cdot n} \in \mathbb{R}^M$ satisfy:
    \begin{enumerate}[label=(\arabic*)]
        \item $\alpha_n \triangleq Y_{\cdot n}^\top (S S^\top)^{-1} S Z_{\cdot n} > 0$;

        \item $\beta_n \triangleq Z_{\cdot n}^\top (I-S^\top (SS^\top)^{-1} S) Z_{\cdot n} > 0$;

        \item $\gamma_n \triangleq Y_{\cdot n}^\top (S S^\top)^{-1} Y_{\cdot n} > 0$.
    \end{enumerate}

\end{assumption}
Recall that in \Cref{sec:problem_formulation} we assume $S$ has full row rank, ensuring that $(SS^\top)^{-1}$ exists. 
Built on the assumption and inspired by recent works on null-space decomposition \cite{nullspace2023, nullspace2025}, we establish the following theorem, which provides the global optimal solution to \eqref{eq:simplified_problem}.

\begin{theorem}
\label{thm:acp}
    Under \Cref{asp:problem}, the problem \eqref{eq:simplified_problem} for each $n=1,2,...,N$ has non-trivial solution
    \begin{equation}
        \label{eq:solution}
        X_{\cdot n}^* = S^\top (S S^\top)^{-1} Y_{\cdot n} + \left(I-S^\top (SS^\top)^{-1} S \right) \frac{\gamma_n}{\alpha_n} Z_{\cdot n}.
    \end{equation}
\end{theorem}

\begin{proof}
Let $S^\dag = S^\top (S S^\top)^{-1}$ denote the Moore-Penrose inverse of $S$.
To facilitate the derivation, we first decompose $X_{\cdot n}$ and $Z_{\cdot n}$ into components lying in the range space $\mathcal{R}(S^\dag) = \{ S^\top (S S^\top)^{-1} \boldsymbol{u} : \boldsymbol{u} \in \mathbb{R}^M \}$ and the null space $\mathcal{N}(S) = \{ \boldsymbol{v} \in \mathbb{R}^C : S\boldsymbol{v} = \mathbf{0} \}$, i.e., the range-null space decomposition:
\begin{align}
    X_{\cdot n} &= S^\top (SS^\top)^{-1} Y_{\cdot n} + \left(I-S^\top (SS^\top)^{-1} S \right) X_{\cdot n} \nonumber \\
    &\triangleq X_{. n, \mathcal{R}} + X_{. n, \mathcal{N}},
\end{align}
\begin{align}
    Z_{\cdot n} &= S^\top (SS^\top)^{-1} S Z_{\cdot n}  + \left(I-S^\top (SS^\top)^{-1} S \right) Z_{\cdot n} \nonumber \\
    &\triangleq Z_{. n, \mathcal{R}} + Z_{. n, \mathcal{N}}.
\end{align}
In this decomposition, $X_{\cdot n, \mathcal{N}}$ is the only optimization variable, whereas $Z_{\cdot n, \mathcal{R}}$, $Z_{\cdot n, \mathcal{N}}$, and $X_{\cdot n, \mathcal{R}}$ are fixed and known.
By the orthogonality between the range space $\mathcal{R}(S^\dag)$ and the null space $\mathcal{N}(S)$, the objective function of problem \eqref{eq:simplified_problem} can be expressed as
\begin{align}
    \frac{X_{\cdot n}^\top Z_{\cdot n}}{\Vert X_{\cdot n}\Vert \Vert Z_{\cdot n} \Vert} &= \frac{1}{\Vert Z_{\cdot n} \Vert} \frac{(X_{\cdot n, \mathcal{R}} + X_{\cdot n, \mathcal{N}})^\top (Z_{\cdot n, \mathcal{R}} + Z_{\cdot n, \mathcal{N}})}{\Vert X_{\cdot n, \mathcal{R}} + X_{\cdot n, \mathcal{N}} \Vert} \nonumber \\
    &= \frac{1}{\Vert Z_{\cdot n} \Vert} \frac{X_{\cdot n, \mathcal{R}}^\top Z_{\cdot n, \mathcal{R}} + X_{\cdot n, \mathcal{N}}^\top Z_{\cdot n, \mathcal{N}}}{\sqrt{\Vert X_{\cdot n, \mathcal{R}} \Vert^2 + \Vert X_{\cdot n, \mathcal{N}} \Vert^2}}.
\end{align}
Therefore, problem \eqref{eq:simplified_problem} can be equivalently reformulated as
\begin{equation}
    \label{eq:decomposed_problem}
    \max_{X_{. n, \mathcal{N}} \in \mathcal{N}(S)} \frac{X_{\cdot n, \mathcal{R}}^\top Z_{\cdot n, \mathcal{R}} + X_{\cdot n, \mathcal{N}}^\top Z_{\cdot n, \mathcal{N}}}{\sqrt{\Vert X_{\cdot n, \mathcal{R}} \Vert^2 + \Vert X_{\cdot n, \mathcal{N}} \Vert^2}}.
\end{equation}
The problem \eqref{eq:decomposed_problem} attains its optimum if and only if $X_{\cdot n, \mathcal{N}}$ and $Z_{\cdot n, \mathcal{N}}$ are collinear, so that their inner product in the numerator is maximized. Hence, let $X_{\cdot n, \mathcal{N}} = \xi Z_{\cdot n, \mathcal{N}}$, $\xi \geq 0$, and the problem can be further simplified to a univariate optimization problem: 
\begin{equation}
    \label{eq:uni_problem}
    \max_{\xi \geq 0} \quad g(\xi),
\end{equation}
where the objective function $g(\xi)$ can be derived as
\begin{align}
    g(\xi) &= \frac{X_{\cdot n, \mathcal{R}}^\top Z_{\cdot n, \mathcal{R}} + \Vert Z_{\cdot n, \mathcal{N}} \Vert^2  \xi}{\sqrt{\Vert X_{\cdot n, \mathcal{R}} \Vert^2 + \Vert Z_{\cdot n, \mathcal{N}} \Vert^2 \xi^2}} \nonumber \\
    &= \frac{\alpha_n + \beta_n \xi}{\sqrt{\gamma_n + \beta_n \xi^2}},
\end{align}
where $\alpha_n$, $\beta_n$, and $\gamma_n$ are defined in \Cref{asp:problem}.
We proceed by differentiating $g(\xi)$ to find the optimal solution. The derivative of $g(\xi)$ is 
\begin{equation}
    g'(\xi) = \frac{\beta_n \gamma_n - \alpha_n \beta_n \xi}{ \left( \beta_n \xi^2 + \gamma_n \right)^{3/2}}.
\end{equation}
By \Cref{asp:problem}, we conclude that $g'(\xi) > 0$ for $\xi \in [0, \gamma_n/\alpha_n)$ and $g'(\xi)<0$ for $\xi \in (\gamma_n/\alpha_n, +\infty)$. 
Thus, $g(\xi)$ takes maximum value at $\xi^* = \gamma_n/\alpha_n$, yielding the optimal solution for problem \eqref{eq:simplified_problem} as
\begin{align}
    X_{\cdot n}^* &= X_{. n, \mathcal{R}} + \xi^* Z_{. n, \mathcal{N}} \nonumber \\
    &= S^\top (S S^\top)^{-1} Y_{\cdot n} + \left(I-S^\top (SS^\top)^{-1} S \right) \frac{\gamma_n}{\alpha_n} Z_{\cdot n}.
\end{align}

\end{proof}

Therefore, for each pixel $n=1,2,\ldots,N$, the optimum upsampled spectral curve $X_{\cdot n}^*$ is obtained as given in \eqref{eq:solution}.
Finally, by concatenating all $X_{\cdot n}^*$, we obtain the final result in matrix form and define the ACP operator as
\begin{equation}
    \label{eq:acp}
    P_{\text{ACP}}(Z, S, Y) \triangleq X^* = S^\dag Y + P_S Z  \Pi,
\end{equation}
where $S^\dag = S^\top (S S^\top)^{-1}$ is the Moore-Penrose inverse of $S$, $P_S = I-S^\top (SS^\top)^{-1} S$ is the orthogonal projection matrix onto null space $\mathcal{N}(S)$, and $\Pi$ is the diagonal matrix of coefficients $\gamma_n/\alpha_n$ given by
\begin{equation}
    \Pi = \begin{pmatrix}
\frac{\gamma_1}{\alpha_1} & & & \\
 & \frac{\gamma_2}{\alpha_2} & & \\
 & & \ddots & \\
 & & & \frac{\gamma_N}{\alpha_N}
\end{pmatrix}.
\end{equation}
Given $Y$, $S$, and $Z$, the upsampled HSI can be computed in closed form using \eqref{eq:acp}, without the need for any iterative optimization or training procedure.
Additionally, \Cref{asp:problem} is generally mild and commonly satisfied in practical scenarios, thus ensuring the broad applicability of the ACP method.

\subsection{Reconstruction Neural Operator}
\label{sec:no}

\begin{figure}[t]
    \centering
    \includegraphics[width=0.9\linewidth]{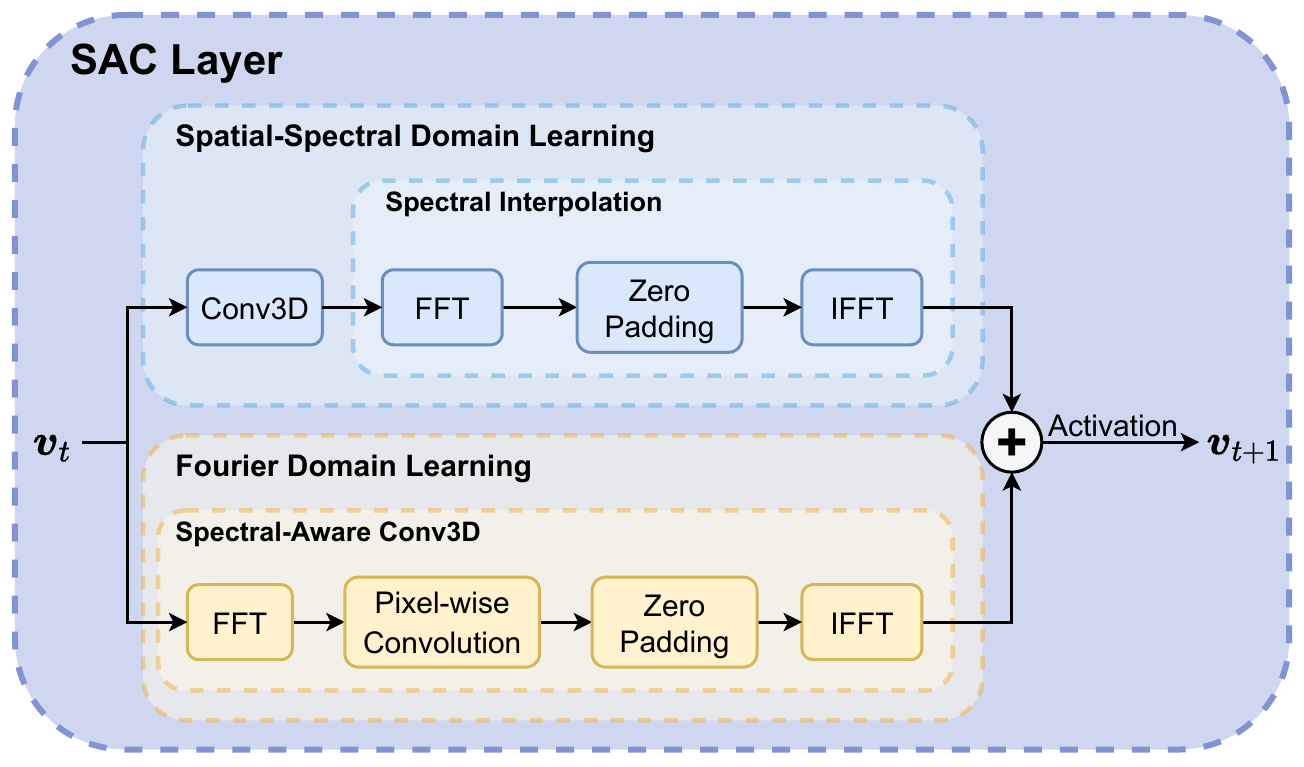}
    \caption{Architecture of the spectral-aware convolution layer. The input $\boldsymbol{v}_t$ goes through two parallel streams: Fourier domain learning and spatial-spectral domain learning. Two outputs are added up, passed through the activation function, and result in the output $v_{t+1}$.} 
    \label{fig:sac_layer}
\end{figure}

In this section, we introduce the neural operator architecture for the reconstruction stage.
In this stage, we aim to learn the continuous mapping between the upsampled HSI $\bar{X}$ and the target HSI $X$ via a data-driven neural network. 
The output $\tilde{X}$ in \eqref{eq:stage2} is expected to be more informative and more accurate than the input $\bar{X}$, i.e., closer to the ground truth $X$.

An overview of the neural operator architecture is provided in \Cref{fig:overview}, with its layer-wise structure shown in \Cref{fig:sac_layer}.
Building on advances in Fourier domain learning for image reconstruction \cite{Ma2024FreqDenoise} and neural operator design \cite{rahman2023uno}, we adopt a spectral-aware convolution module and integrate a U-shaped structure to enable multi-scale extraction of spatial and spectral features.
The SAC module performs Fourier transforms across both spectral and spatial dimensions, while restricting the pixel-wise convolution to the spectral domain using a fixed set of Fourier modes.
The algorithm of the spectral-aware convolution is shown in \Cref{alg:saconv3d}. 

\begin{algorithm}[t]
\caption{Spectral-Aware Convolution} \label{alg:saconv3d}
\begin{algorithmic}[1]
\REQUIRE Input tensor $x:[B, d_{\text{in}}, H, W, C]$, weight $W:[d_{\text{in}},d_{\text{out}},d_{\text{modes}}]$.

\STATE Transform $x$ to Fourier domain:
\STATE $\tilde{x} \gets \text{rfft}(x)$ \hfill \# shape: $[B, d_{\text{in}}, H, W, C//2+1]$

\STATE Initialize output in frequency domain: 
\STATE $\tilde{y} \gets 0$ \hfill \# shape: $[B, d_{\text{out}}, H, W, C//2+1]$

\STATE Pixel-wise convolution:
\STATE $\tilde{y}[..., :d_{\text{modes}}] \gets \text{einsum}($
\STATE \hspace*{1.5em}$\text{``bixyz,ioz} \to \text{boxyz"},$ \\
\STATE \hspace*{1.5em}$\tilde{x}[..., :d_{\text{modes}}], W$
\STATE $)$

\STATE Apply inverse Fourier transform:
\STATE $y \gets \text{irfft}(\tilde{y})$
\RETURN $y$
\end{algorithmic}
\end{algorithm}

In general, based on the formulation of the neural operator in \Cref{sec:operator_learning}, the structure of the reconstruction neural operator $R_{no}(\cdot, \boldsymbol{w})$ can be summarized as follows.
\begin{enumerate}
    \item \textbf{Lifting:} The upsampled HSI $\bar{X}$, together with the input coordinate $\boldsymbol{w}$, is lifted into a higher-dimensional feature space through a multilayer perceptron (MLP), formulated as
    \begin{equation}
        \tilde{X}_0 = \text{MLP}_{P}\left(\text{Concat}\left(\bar{X}, \boldsymbol{w} \right)\right),
    \end{equation}
    where $\tilde{X}_0$ is the first hidden representation, $\text{Concat}(\cdot)$ is the concatenation operation, and trivial reshaping and expanding operations are omitted for clarity.

    \item \textbf{U-shaped Kernel Integration:} We iteratively update the hidden representation through several U-shaped SAC layers to yield the next representation, as illustrated in \Cref{fig:overview} and \Cref{fig:sac_layer}. 
    To formulate this, let $T_c$ be the number of layers in the contracting path and expansive path, and $T_r$ be the number of layers in the transformation path. The iterative updates in the contracting path and transformation path can be formulated as
    \begin{equation}
        \tilde{X}_{t+1} = \sigma\left( \mathcal{L}_t\left( \tilde{X}_t \right) + \text{SAC}_t\left(\tilde{X}_t\right) \right),
    \end{equation}
    where $t=0, 1, \dots, T_c-1, T_c, \dots, T_c+T_r-1$, $\mathcal{L}_t(\cdot)$ and $\text{SAC}_t(\cdot)$ represent the linear convolution operator in the spatial-spectral domain and the spectral-aware convolution operator in the Fourier domain at $t$ step respectively, and $\sigma(\cdot)$ is the activation function. While for the expansive path, the iterative updates can be expressed as 
    \begin{align}
        \tilde{X}_{t+1} = \sigma \left( \mathcal{S}_t\left( \text{Concat}\left( \tilde{X}_t, \tilde{X}_{2T_c + T_r - 1-t} \right) \right) \right),
    \end{align}
    where $t=T_c+T_r, \dots, 2T_c + T_r - 1$, $\mathcal{S}_t(\cdot) = \mathcal{L}_t\left( \cdot \right) + \text{SAC}_t\left( \cdot \right)$.

    \item \textbf{Projection:} We finally map the last hidden representation $\tilde{X}_{2T_c + T_r}$ to the estimated HSI $\tilde{X}$ via another MLP, that is,
    \begin{equation}
        \tilde{X} = \text{MLP}_{Q}\left( \tilde{X}_{2T_c + T_r} \right).
    \end{equation}
\end{enumerate}

\section{Experiments}

\begin{figure*}[!t]
    \centering
    \includegraphics[width=0.96\textwidth]{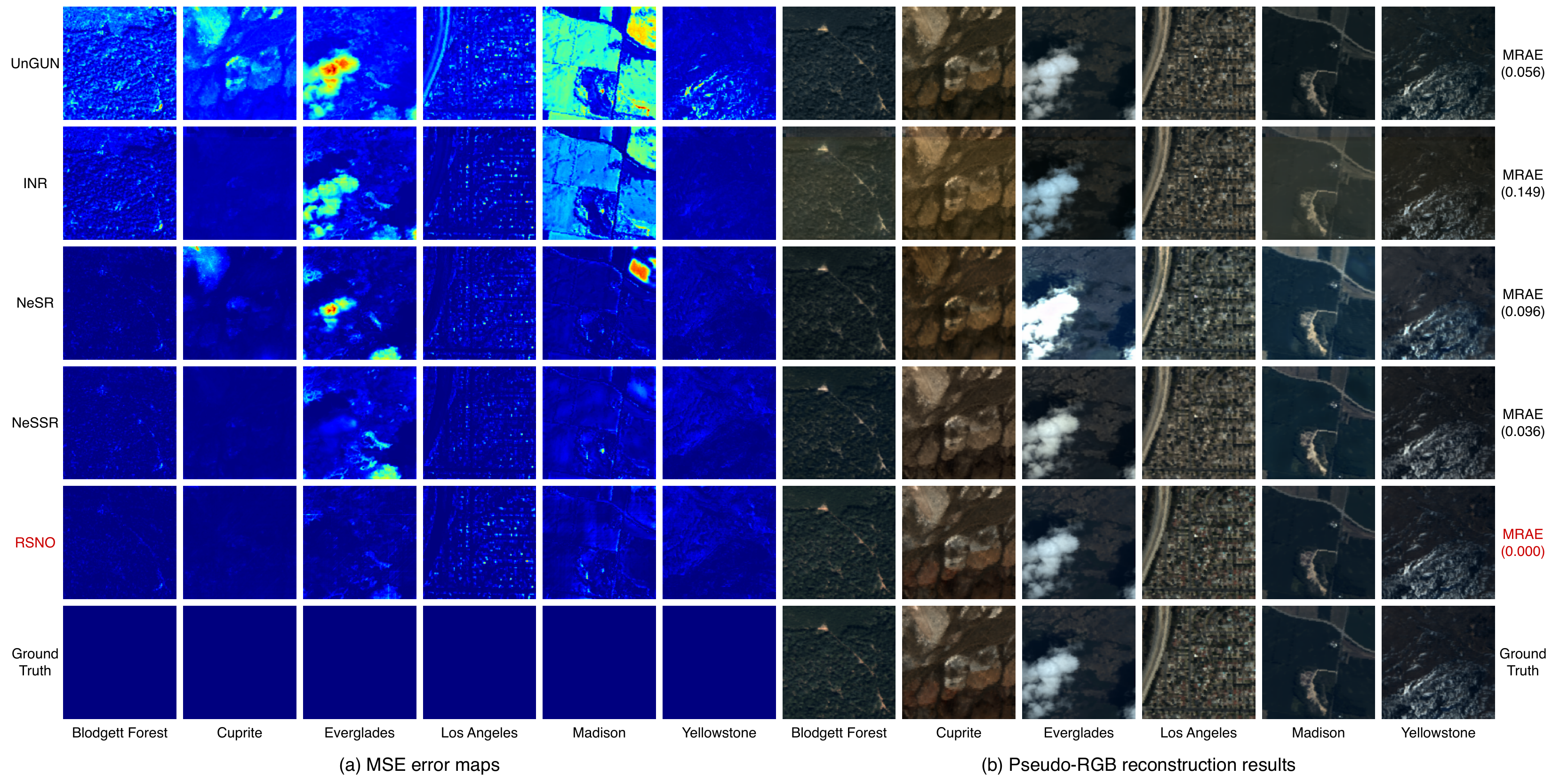}
    \caption{Visual comparison of reconstruction results across different sites. Each row corresponds to a spectral reconstruction method, while each column corresponds to a scene from a specific site. In (b), the MRAE reconstruction error is shown on top of each image.}
    \label{fig:visual_comparison}
\end{figure*}

In this section, we present numerical experiment results to validate our approach under various settings. 
We begin by introducing the datasets in \Cref{sec:datasets} and outlining the basic experimental setups in \Cref{sec:setups}.
We first evaluate our model under the conventional discrete SSR setting, with results reported in \Cref{sec:comparative}. 
We then assess its performance in continuous spectral super-resolution in \Cref{sec:assr}. 
Finally, an ablation study is presented in \Cref{sec:ablation} to analyze how and why the proposed design is effective.

\subsection{Datasets}
\label{sec:datasets}
We collect data from the NASA Jet Propulsion Laboratory (JPL), originally acquired by the widely recognized AVIRIS (Airborne Visible/Infrared Imaging Spectrometer) sensors, and curate it into a new dataset for our study\footnote{Data available at https://aviris.jpl.nasa.gov/dataportal}.
The dataset comprises hyperspectral images spanning 224 contiguous spectral bands, ranging from 400 to 2500 nm.
Specifically, we select various typical and representative sites, each comprising multiple scenes.
The scene characteristics, such as season and spatial resolution, are consistent within each site but differ significantly across different sites.
The summary of sites used in our experiment is provided in \Cref{tab:datasets}.

\begin{table}[htbp]
\centering
\caption{AVIRIS Sites Selected for the Experiments}
\label{tab:datasets}
\renewcommand{\arraystretch}{1.2}
\begin{tabular}{c||ccc}
\hline\hline
Site Name & Land Cover Type & Flight ID & Resolution (m) \\
\hline
Blodgett Forest & Forest & f130803t01 & 3.6 \\
Cuprite & Desert & f060502t01 & 3.4 \\
Everglades & Wetland & f100523t01 & 17 \\
Los Angeles & Urban area & f130821t01 & 3.5 \\
Madison & Rural area & f110816t01 & 7.4 \\
Yellowstone & Mountainous area & f060925t01 & 15.1 \\
\hline \hline
\end{tabular}
\end{table}

\subsection{Implementation Details}
\label{sec:setups}
\subsubsection{Basic Setup}
In our experiments, we use the direct beam radiation predicted by SMARTS \cite{gueymard1995smarts2} based on the 1976 U.S. Standard Atmosphere as the radiative spectral prior, which serves as the angular guidance $Z$ in \eqref{eq:stage1}.
For the SRF, we utilize the database provided by \cite{jiang2013space}, which includes 28 camera sensitivity functions covering a variety of types.
We generate the paired RGB image by multiplying the normalized SRF randomly selected from the database and the HSI from the AVIRIS dataset we collected in \Cref{sec:datasets}, as shown in \eqref{eq:Y=SX}.
For hyperparameters of the neural operator, we set maximum Fourier modes $d_{\text{modes}} = 16$, number of hidden channels $d=32$, and number of contracting and transformation layers $T_c=T_r=4$. 
Coordinate-based terms, including SRF $S_{\boldsymbol{w}}$ and radiative prior $Z_{\boldsymbol{w}}$, are interpolated at wavelength $\boldsymbol{w}$ via kernel regression.

\subsubsection{Loss Function}
In our experiments, we construct the loss function of our model by adding up the $L1$ and the SAM criterion, which is
\begin{equation}
    \label{eq:loss}
    \mathcal{L}(\hat{X}, X) = L_1(\hat{X}, X) + \lambda \text{SAM}(\hat{X}, X),
\end{equation}
where $\lambda$ is the trade-off parameter (we set $\lambda=0.1$ in our experiments), $\text{SAM}(\cdot, \cdot)$ is defined in \eqref{eq:sam}, and $L_1(\cdot, \cdot)$ is calculated by
\begin{equation}
    \label{eq:l1}
    L_1(\hat{X}, X) = \frac{1}{CN} \sum_{c,n} \vert \hat{X}_{cn} - X_{cn} \vert.
\end{equation}

\subsubsection{Metrics}
We adopt four commonly used metrics to evaluate the reconstruction performance: the Mean Relative Absolute Error (MRAE), the Peak Signal-to-Noise Ratio (PSNR), the Structural Similarity Index Measure (SSIM), and the SAM. 
Lower MRAE and SAM values, together with higher PSNR and SSIM, indicate better reconstruction quality.

\subsection{Discrete Spectral Super-Resolution}
\label{sec:comparative}

\begin{figure*}[!t]
    \centering
    \subfloat[Blodgett Forest]{\includegraphics[width=0.32\textwidth]{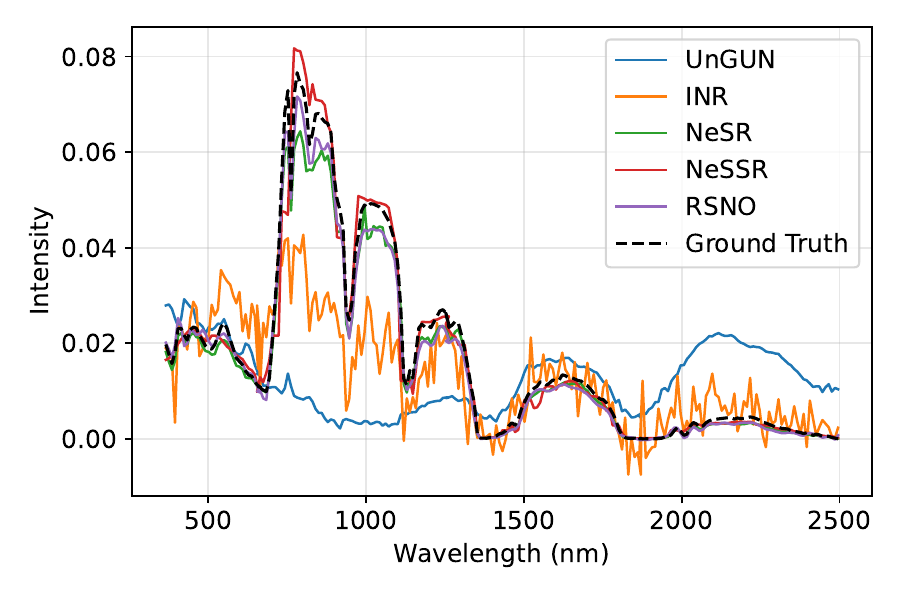}}
    \hfill
    \subfloat[Cuprite]{\includegraphics[width=0.32\textwidth]{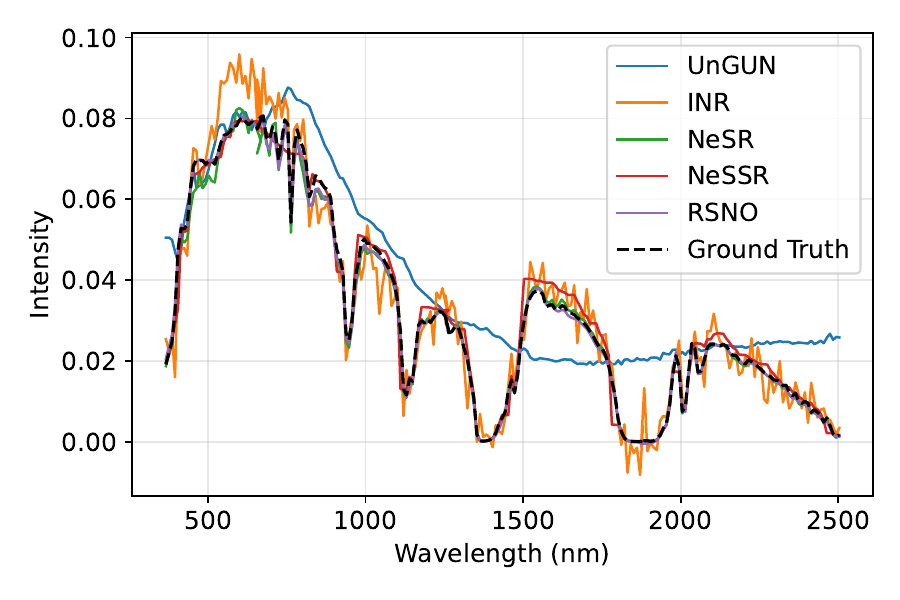}}
    \hfill
    \subfloat[Everglades]{\includegraphics[width=0.32\textwidth]{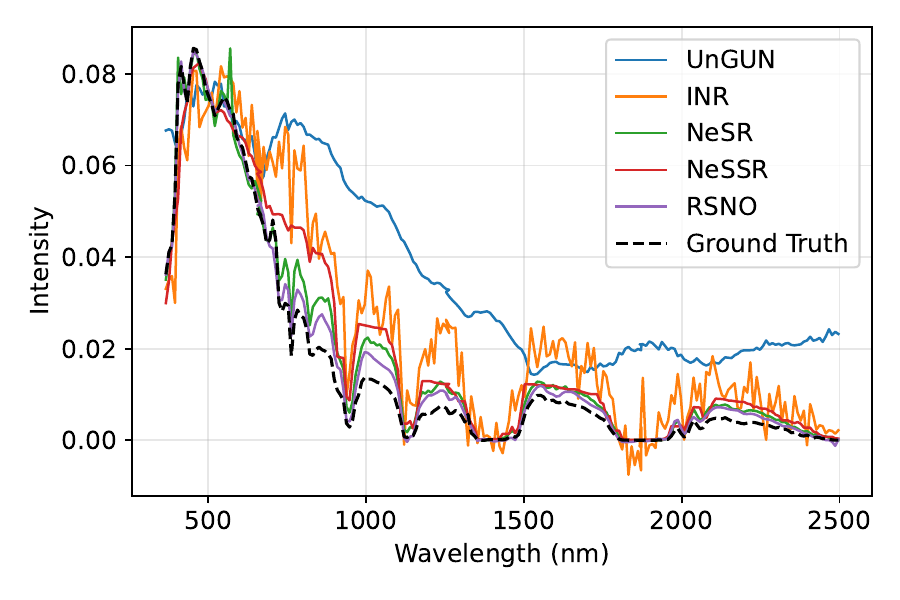}}\\
    \vspace{1mm}
    \subfloat[Los Angeles]{\includegraphics[width=0.32\textwidth]{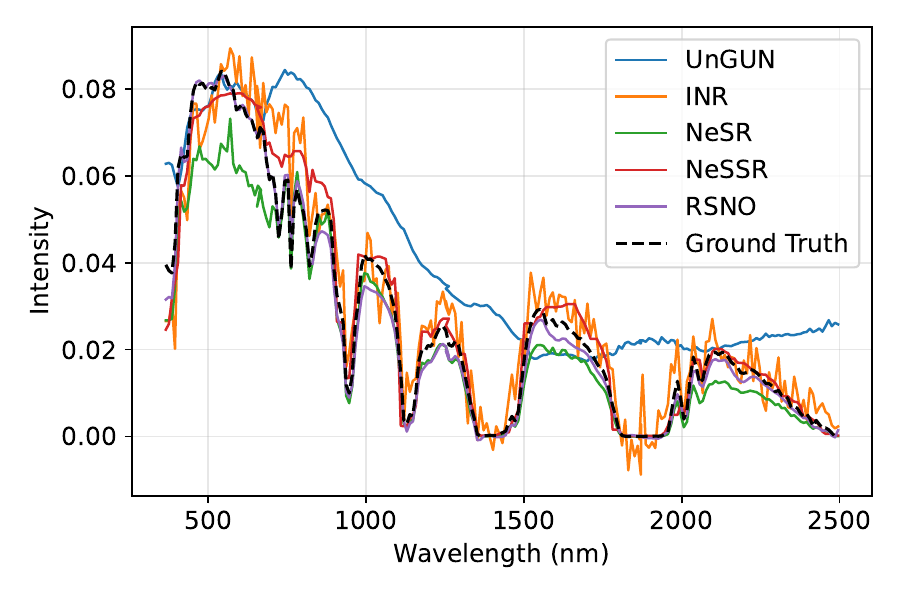}}
    \hfill
    \subfloat[Madison]{\includegraphics[width=0.32\textwidth]{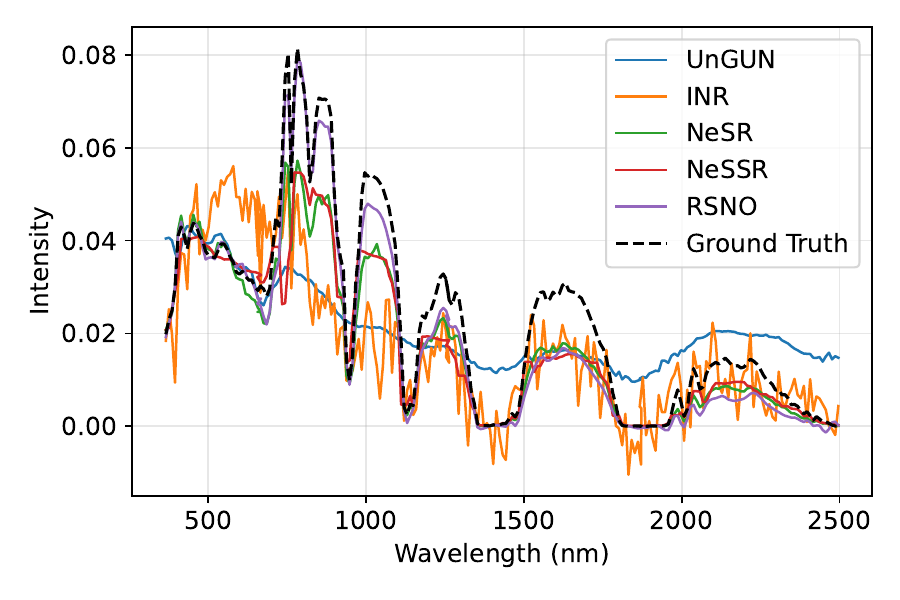}}
    \hfill
    \subfloat[Yellowstone]{\includegraphics[width=0.32\textwidth]{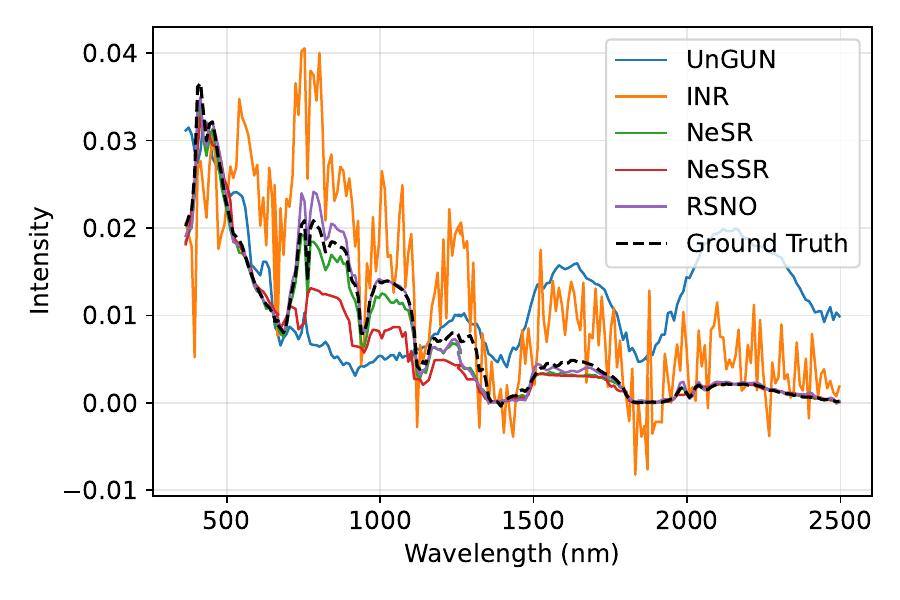}}
    
    \caption{Comparison of predicted spectral signatures for representative reconstructed pixels from six benchmark sites (Blodgett Forest, Cuprite, Everglades, Los Angeles, Madison, and Yellowstone), reconstructed using different methods.}
    
    \label{fig:spectra_comparison}
\end{figure*}

We conduct comparative experiments in the conventional discrete SSR setting to evaluate the performance of the proposed RSNO against several representative methods, including UnGUN~\cite{qu2023unmixing}, INR~\cite{Kaiwei2023INR-HSISR}, NeSR~\cite{xu2022nesr}, and NeSSR~\cite{xu2025nessr}. 
All models are evaluated under the same training-testing protocol.

Table~\ref{tab:compare_overall} reports the overall quantitative results on the test datasets. 
Among all baselines, our RSNO consistently achieves the best performance, with the lowest MRAE and SAM, the highest PSNR and SSIM, and the fewest parameters among high-performing methods.
These results highlight the efficiency and effectiveness of our design.
Besides, we conduct per-site evaluations to further compare the performance of the four methods. 
The results are summarized in \Cref{tab:compare_six}. 
RSNO consistently delivers top or near-top performance in most scenarios, particularly excelling in complex scenes such as desert and wetland. 
However, results in the urban area and mountainous area show that RSNO achieves slightly inferior performance compared with the competing methods.

We also conduct a visual comparison among all models, as shown in \Cref{fig:visual_comparison}.
From error maps we can observe that RSNO achieves lower MSE in most scenarios, except for \textit{Los Angeles}, which is the most complex scene with dense man-made structures.
Pseudo-RGB reconstruction results demonstrate that our method is capable of faithfully restoring RGB images, whereas other methods exhibit visual artifacts or unnatural color distortions.
This advantage mainly stems from the refinement step, which explicitly enforces zero reconstruction loss with ACP, ensuring accurate color consistency.
Additionally, we plot the predicted spectrum on six sites, which are shown in \Cref{fig:spectra_comparison}.
Our method yields more natural and physically plausible spectra that better align with the ground truth, especially in wavelength regions affected by atmospheric absorption. 
In contrast, other approaches tend to either overfit noisy variations or underfit the true reflectance shape.

In general, both quantitative and qualitative comparisons indicate that RSNO achieves state-of-the-art performance and more stable results in the discrete setting across varied geographical contexts with fewer parameters.
However, certain limitations are also observed. 
In particular, RSNO may suffer from less accurate predictions in complex scenarios such as the urban area, where highly intricate man-made structures pose significant challenges. 
Addressing this issue remains an important direction for future work.

\begin{table*}[htbp]
\centering
\caption{Comparative Results on the Overall Datasets}
\label{tab:compare_overall}
\setlength{\tabcolsep}{6pt} 
\renewcommand{\arraystretch}{1.2} 
\begin{tabular}{c|c|c||ccccc}
\hline\hline
Method & Params (M) & Continuous & MRAE $\downarrow$ & PSNR $\uparrow$ & SAM $\downarrow$ & SSIM $\uparrow$ \\
\hline
UnGUN \cite{qu2023unmixing} & 1.1 & \ding{55} & 0.611 & 35.167 & 0.536 & 0.808 \\
INR\cite{Kaiwei2023INR-HSISR} & 63.9 & \ding{55} & 0.366 & 39.592 & 0.312 & 0.906 \\
NeSR\cite{xu2022nesr} & 29.3 & \checkmark & 0.190 & 42.885 & \underline{0.142} & 0.954 \\
NeSSR\cite{xu2025nessr} & 5.3 & \checkmark & \underline{0.186} & \underline{43.367} & 0.178 & \underline{0.963} \\
\rowcolor{mycolor}
RSNO (Ours)  & 4.8 & \checkmark & \textbf{0.160} & \textbf{45.400} & \textbf{0.137} & \textbf{0.973} \\
\hline\hline
\end{tabular}

\end{table*}

\begin{table*}[htbp]
\centering
\caption{Comparative Results for Each Site}
\label{tab:compare_six}
\renewcommand{\arraystretch}{1.2}

\begin{tabular}{c||cc|cc|cc|cc|cc|cc}
\hline\hline
\multirow{3}{*}{Method}  & 
\multicolumn{2}{c|}{Blodgett Forest} &
\multicolumn{2}{c|}{Cuprite} &
\multicolumn{2}{c|}{Everglades} &
\multicolumn{2}{c|}{Los Angeles} &
\multicolumn{2}{c|}{Madison} &
\multicolumn{2}{c}{Yellowstone} \\
 & 
\multicolumn{2}{c|}{\textit{Forest}} &
\multicolumn{2}{c|}{\textit{Desert}} &
\multicolumn{2}{c|}{\textit{Wetland}} &
\multicolumn{2}{c|}{\textit{Urban area}} &
\multicolumn{2}{c|}{\textit{Rural area}} &
\multicolumn{2}{c}{\textit{Mountainous area}} \\
\cline{2-13}
 & PSNR $\uparrow$ & SAM $\downarrow$ 
 & PSNR $\uparrow$ & SAM $\downarrow$ 
 & PSNR $\uparrow$ & SAM $\downarrow$ 
 & PSNR $\uparrow$ & SAM $\downarrow$ 
 & PSNR $\uparrow$ & SAM $\downarrow$ 
 & PSNR $\uparrow$ & SAM $\downarrow$ \\
\hline

UnGUN \cite{qu2023unmixing} & 
32.537 & 1.001 &
36.546 & 0.310 &
33.005 & 0.455 &
34.616 & 0.454 &
31.402 & 0.805 &
39.468 & 0.578 \\

INR \cite{Kaiwei2023INR-HSISR} & 
35.984 & 0.483 &
44.403 & 0.130 &
35.584 & 0.331 &
39.402 & 0.254 &
33.313 & 0.497 &
42.553 & 0.390 \\

NeSR \cite{xu2022nesr} & 
\underline{42.910} & \textbf{0.106} &
48.152 & \underline{0.064} &
36.803 & \underline{0.199} &
40.186 & \underline{0.150} &
38.686 & \underline{0.181} &
\textbf{46.416} & \textbf{0.184} \\

NeSSR \cite{xu2025nessr} & 
41.353 & 0.193 &
\underline{48.489} & 0.086 &
\underline{39.324} & 0.240 &
\textbf{42.376} & 0.164 &
\underline{38.856} & 0.208 &
44.968 & 0.242 \\
\rowcolor{mycolor}
RSNO (Ours) & 
\textbf{43.296} & \underline{0.112} &
\textbf{53.423} & \textbf{0.049} &
\textbf{42.188} & \textbf{0.183} &
\underline{41.884} & \textbf{0.148} &
\textbf{41.601} & \textbf{0.125} &
\underline{45.575} & \underline{0.221} \\

\hline\hline
\end{tabular}
\end{table*}

\begin{table*}[ht]
\centering
\caption{Quantitative Evaluation of Continuous Spectral Reconstruction}
\label{tab:cont_zero}
\setlength{\tabcolsep}{5pt}
\renewcommand{\arraystretch}{1.2}
\begin{tabular}{c||cccc|cccc}
\hline\hline
\multirow{2}{*}{Method} &  
\multicolumn{4}{c|}{$2 \times$} & 
\multicolumn{4}{c}{$4 \times$} \\
\cline{2-9} 
& MRAE $\downarrow$ & PSNR $\uparrow$ & SAM $\downarrow$ & SSIM $\uparrow$
& MRAE $\downarrow$ & PSNR $\uparrow$ & SAM $\downarrow$ & SSIM $\uparrow$ \\
\hline
NeSR \cite{xu2022nesr} 
& 0.565 & 34.116 & 0.374 & 0.681
& 0.563 & 33.920 & 0.309 & 0.677 \\
NeSSR \cite{xu2025nessr} 
& 0.359 & 38.130 & 0.323 & 0.849
& 0.436 & 35.441 & 0.431 & 0.788 \\
\rowcolor{mycolor}
RSNO (Ours) 
& \textbf{0.205} & \textbf{43.629} & \textbf{0.176} & \textbf{0.956}
& \textbf{0.254} & \textbf{41.865} & \textbf{0.224} & \textbf{0.933} \\
\hline\hline
\end{tabular}
\end{table*}

\subsection{Continuous Spectral Super-Resolution}
\label{sec:assr}

To validate the effectiveness of our continuous modeling and evaluate the generalization and flexibility of our method, we conduct experiments on continuous spectral reconstruction.
We train the model using only a small number of uniformly downsampled spectral bands and test it on the full spectral resolution. 
In our experiments, we apply $2 \times$ and $4 \times$ downsampling for continuous spectral reconstruction.
We set $d_{\text{modes}}=12$ in this experiment to avoid $d_{\text{modes}}>C//2+1$ for high downsampling scales.
We compare our method against two other continuous methods: NeSR and NeSSR.
Table~\ref{tab:cont_zero} summarizes the quantitative results. 
Our method achieves best performance in both settings, indicating strong generalization to sparsely-sampled spectral bands.

The good performance of RSNO stems from two factors: the neural operator backbone ensuring resolution-invariant spectral interpolation, and the ACP method guided by radiative spectral prior ensuring radiative-structured spectral shape from seen to unseen bands.
These results highlight RSNO’s robustness and generalization ability, making it suitable for practical hyperspectral applications.

\subsection{Ablation Study}
\label{sec:ablation}

In this section, we conduct ablation studies to further investigate the effectiveness of the proposed method.
The key innovation of our method lies in the ACP, which integrates physical priors into a data-driven pipeline during the initial upsampling and enforces hard constraints during final refinement.
To validate its effectiveness, we perform ablation studies on the upsampling stage and the refinement stage. 
For the upsampling stage, we use a zero matrix as the angular guidance $Z$ to obtain the upsampled HSI, rather than using the radiative prior, which is also equivalent to directly employing the Moore-Penrose inverse.
To compare hard and soft constraints, we directly remove the final refinement stage and add an extra error term on the loss function as the regularization to enforce low reconstruction error, which is
\begin{equation}
    \label{eq:loss_regular}
    \mathcal{L}_r(\hat{X}, X) = \mathcal{L}(\hat{X}, X) + \alpha L_1(S_{\boldsymbol{w}}\hat{X}, S_{\boldsymbol{w}}X),
\end{equation}
where $\alpha$ is the trade-off parameter, and we set $\alpha = 0.5$ in the ablation experiment.
The results are listed in \Cref{tab:ablation}. The results indicate that both the radiative prior and the final refinement play critical roles in overall performance, thereby demonstrating the effectiveness of the ACP method.

\begin{table}
\centering
\caption{Ablation Study of Radiative Prior and Refinement}
\label{tab:ablation}

\renewcommand{\arraystretch}{1.2}

\begin{tabular}{c|c||cc}
\hline\hline
Radiative Prior & Refinement & PSNR $\uparrow$ & SAM $\downarrow$ \\
\hline
\checkmark & \checkmark & \textbf{45.400} & \textbf{0.137} \\
\ding{55}  & \checkmark & 43.118 & 0.167 \\
\checkmark & \ding{55}  & 40.163 & 0.170 \\
\ding{55}  & \ding{55}  & 38.760 & 0.178 \\
\hline\hline
\end{tabular}
\end{table}

\section{Conclusion}
In this article, we propose a novel physics-informed approach for continuous spectral reconstruction. 
By continuously reformulating the problem and embedding physical priors within a data-driven framework, our method yields more realistic reconstructions and effectively balances interpretability and adaptability. 
Experiments across diverse real-world scenes and problem settings indicate better reconstruction accuracy and robustness, especially in the continuous setting. 
This work highlights the potential of continuous spectral modeling and the integration of domain knowledge with deep learning to advance spectral super-resolution for practical applications.

\bibliography{reference}
\bibliographystyle{IEEEtran}

\end{document}